\documentclass[%
 reprint,
superscriptaddress,
nofootinbib,
amsmath,amssymb,
aps,
prl,
]{revtex4-2}

\usepackage{graphicx}
\usepackage{dcolumn}
\usepackage{bm}
\usepackage{blindtext}
\usepackage{xcolor}
\usepackage{amsthm}
\usepackage{amsmath}
\usepackage{amssymb}
\usepackage{multirow}
\usepackage{multibib}
\usepackage{hyperref}

\newtheorem*{remark}{Remark}
\newtheorem{theorem}{Theorem}
\newcommand{\tr}{{\rm tr}}
\newcites{e}{Additional References}

\begin{document}

\preprint{DESY 22-}

\title{Learning Trivializing Gradient Flows for Lattice Gauge Theories}

\author{Simone Bacchio}
\affiliation{
 Computation-based Science and Technology Research Center, The Cyprus Institute, Nicosia, Cyprus
}%
\author{Pan Kessel}
\affiliation{Machine Learning Group, Technische Universit\"{a}t Berlin, Berlin, Germany}
\affiliation{
BIFOLD—Berlin Institute for the Foundations of Learning and Data, Berlin, Germany
}
\author{Stefan Schaefer}
\affiliation{John von Neumann-Institut f\"ur Computing NIC, Deutsches Elektronen-Synchrotron DESY, Germany
}
\author{Lorenz Vaitl}
\affiliation{Machine Learning Group, Technische Universit\"{a}t Berlin, Berlin, Germany}

\date{\today}

\begin{abstract}
We propose a unifying approach that starts from the perturbative construction of trivializing maps by L\"uscher and then improves on it by learning. The resulting continuous normalizing flow model can be implemented using common tools of lattice field theory and requires several orders of magnitude fewer parameters than any existing machine learning approach. Specifically, our model can achieve competitive performance with as few as 14 parameters while existing deep-learning models have around 1 million parameters for $SU(3)$ Yang--Mills theory on a $16^2$ lattice. This has obvious consequences for training speed and interpretability. It also provides a plausible path for scaling machine-learning approaches toward realistic theories.
\end{abstract}
\maketitle

\paragraph{Introduction.}
Critical slowing down constitutes one of the fundamental challenges of modern computational sciences. A particular situation in which critical slowing down is observed is the continuum limit of lattice field theories. An important theme of research of the last few decades is to construct algorithms, such as cluster methods \cite{wolff1989collective, swendsen1987nonuniversal}, that are (relatively) insensitive to critical slowing down. Despite significant effort, a cluster algorithm for lattice quantum chromodynamics has unfortunately not been found. 

An interesting alternative approach to circumvent critical slowing down was proposed in a seminal paper \cite{luscher2010trivializing} by L\"uscher more than ten years ago: building on previous work in supersymmetric field theories \cite{nicolai1980supersymmetry}, L\"uscher proposed to use a field redefinition that maps pure $SU(3)$ Yang-Mills theory to its strong coupling limit and therefore trivializes it. In the redefined field variables, the theory is not (severely) affected by critical slowing down even if standard HMC algorithms are applied. Despite the considerable conceptual appeal, the main practical challenge of this approach is to construct such trivializing maps. L\"uscher proposed a perturbative construction of a gradient flow that realizes the trivializing map to linear order in the flow time. Unfortunately, the expansion is around the strong coupling limit and therefore the approximation deteriorates as the continuum is approached. Indeed, in HMC simulations of the  $\mathrm{CP}^{N-1}$ model the autocorrelations were reduced, but the scaling was not substantially changed~\cite{Engel:2011re}.

Recently, trivializing maps have gained renewed attention since they can be combined with deep learning \cite{albergo2019flow, kanwar2020equivariant, boyda2021sampling, albergo2021flow, hackett2021flow, albergo2022flow, abbott2022gauge, abbott2022sampling, del2021efficient, del2021machine, nicoli2021machine, nicoli2021estimation, de2021scaling, gerdes2022learning, finkenrath2022tackling, abbott2022sampling, Caselle:2022acb, Favoni:2022mcg, Favoni:2020reg}. In this new line of work, the trivializing map is modeled by a deep neural network. The parameters of the neural network are then trained to trivialize the redefined theory. This approach is interesting for at least three reasons: i.) it circumvents any expansion in flow time, ii.) it is (at least in principle) completely general as neural networks are universal approximators, and iii.) it establishes a connection to the rapidly progressing field of machine learning raising the exciting prospect of using specialized deep learning silicon, such as Tensor Processing Units (TPUs), for lattice simulations. Despite these advantages, state-of-the-art methods for learning trivializing maps only work for low-dimensional theories. This is partly because the training relies on self-sampling from the model. In order to attain a useful gradient signal, the model has to probe relevant regions of field space of the lattice field theory. As the dimensionality of the theory increases, these regions are very unlikely to be sampled and training will fail. This is one of the reasons for the poor volume scaling of current deep-learning-based approaches \cite{abbott2022aspects, del2021efficient, del2021machine, finkenrath2022tackling}.

In this work, we outline a strategy that unifies L\"uscher's perturbative with the recent machine learning approach, in particular continuous normalizing flows \cite{grathwohl2018ffjord, kohler2020equivariant, de2021scaling, gerdes2022learning}. More specifically, we propose to use the same gradient flow as L\"uscher but instead of fixing its coefficients by perturbation theory to learn them by machine learning techniques, namely the adjoint state method. This has the advantage that we can use L\"uscher's perturbative results to initialize the machine learning model. The resulting model is not only manifestly equivariant under all global as well as local symmetries of the theory, but it is also constrained to a low number of required free parameters. We demonstrate that we can obtain comparable performance to the current state-of-the-art deep learning models, which have around 1 million parameters, with as few as $\mathcal{O}(10)$ parameters, and significantly outperform them by using $\mathcal{O}(100)$ parameters. We demonstrate that the low number of parameters and the perturbative initialization are particularly beneficial in the early stage of training. This is encouraging as this phase is the main hurdle in scaling machine learning approaches. A further advantage of our method is that it can be implemented using standard lattice tools and does not require any specialized deep learning libraries. 

The main objective of this paper is to introduce our approach and demonstrate its superior performance when compared to previous methods. For this, we will restrict ourselves to two-dimensional Yang--Mills theory as this is the benchmark on which we can compare to the existing literature leaving applications to higher-dimensional theories for future work. 
\\

\paragraph{Trivializing Maps.}
In lattice gauge theory, the expectation value of a physical observable $\mathcal{O}$ is given by the Wick-rotated path integral
\begin{align}
    \langle \mathcal{O} \rangle = \frac{1}{\mathcal{Z}} \int \textrm{D}[U] \, \mathcal{O}(U) \, \exp(-S(U))
\end{align}
discretized on a lattice $\Lambda$.
Using a (diffeomorphic) field redefinition $U = \mathcal{F}(V)$, this expectation value can be rewritten as
\begin{align}
    \langle \mathcal{O} \rangle =\frac{1}{\mathcal{Z}} \int \textrm{D}[V] \, \mathcal{O}(\mathcal{F}(V)) \, \exp(-S_{\mathcal{F}}(V)) \,, \label{eq:pullback_expectation}
\end{align}
where we have defined
\begin{align}
    S_{\mathcal{F}}(V)=S(\mathcal{F}(V)) - \ln \textrm{det} \, \mathcal{F}_*(V) \,.\label{eq:zeroflowtimeaction}
\end{align}
The last term involving the Jacobian $\mathcal{F}_*$ is due to the change of measure $\textrm{D}[U] = \textrm{D}[V] \det \mathcal{F}_{*}(V)$. 

For a trivializing map $\mathcal{F}$, this measure contribution cancels the action up to a possible constant, i.e.
\begin{align}
    S_{\mathcal{F}}(V)= \textrm{const.} \,.\label{eq:trivializing_condition}
\end{align}
The expectation value \eqref{eq:pullback_expectation} can thus be calculated using the uniform density.
Such trivializing maps $\mathcal{F}$ can be constructed analytically for certain supersymmetric field theories \cite{nicolai1980supersymmetry}. For the case of $SU(N)$ Yang--Mills theory, a perturbative construction was put forward by L\"uscher in \cite{luscher2010trivializing}. 

Recently, it has been proposed to use machine learning to obtain trivializing maps non-perturbatively. In this approach, the redefinition $\mathcal{F}_\theta$ is given by a bijective machine learning model with parameters $\theta$ \cite{albergo2019flow, kanwar2020equivariant, boyda2021sampling, albergo2021flow, hackett2021flow, albergo2022flow, abbott2022gauge, abbott2022sampling, del2021efficient, del2021machine, nicoli2021machine, nicoli2021estimation, de2021scaling, gerdes2022learning, finkenrath2022tackling}. The model is then trained by minimizing the objective function
\begin{align}
    \mathcal{C}(\theta) = \langle S_{\mathcal{F}_\theta}(V) \rangle \,, \label{eq:kl}
\end{align}
using stochastic gradient descent. 
The parameters $\hat{\theta}$ are a global minimum of the objective $\mathcal{C}(\theta)$ if and only if the corresponding map $\mathcal{F}_{\hat{\theta}}$ is trivializing, i.e. fulfills the trivializing condition \eqref{eq:trivializing_condition}. We refer to the Supplementary Material for a proof.

In practice, we cannot expect the model to be perfectly trained, i.e. $\mathcal{F}_\theta$ does not fulfill the trivializing condition \eqref{eq:trivializing_condition} and thus does not completely reduce the target density
$p(V) = \frac{1}{\mathcal{Z}} \exp(-S_{\mathcal{F}_\theta}(V))$ to the uniform density. One can, however, use the uniform density, $q(V)={\rm const}$, as a proposal for a Markov chain to sample from $p(V)$. Specifically, one advances the Markov chain from a previous configuration $V$ to some current configuration by accepting a candidate $V' \sim q$ with probability
\begin{align}\label{eq:NMCMC}
    p_A = \min\left(1, \frac{w(V')}{w(V)} \right) \,,
\end{align}
with the importance weight $w(V) = \frac{p(V)}{q(V)}$. For a sufficiently trained model, the proposal $V'$ for the update will be accepted with high probability. As a result, autocorrelation will be low as it can only arise due to repeated rejection of proposals (since the proposals are sampled independently from $q$). 
\\

\paragraph{L\"uscher's Perturbative Approach.}
L\"uscher proposed a flow equation given by
\begin{align}
    \dot{U}_t = Z_t (U_t) \, U_t \label{eq:ode}
\end{align}
which is generated by an algebra-valued link field $[Z_t(U)](x, \mu) \in \mathfrak{su}(N)$.\footnote{$[\dot{U}_tU_t^\dagger](x, \mu)$ is obviously anti-hermitian and can easily be shown to be traceless by using Jacobi's formula. This in turn implies that $Z_t$ is Lie-algebra valued.} If $Z_t$ is a smooth function, the solution $U_t$ is unique for a given initial condition $U_0=V$ at any $t \in \mathbb{R}$. Therefore, the flow equation implicitly defines a bijective field redefinition $\mathcal{F}(V)=U_t$, where we suppress the dependency of the map $\mathcal{F}$ on the chosen time $t$ for notational simplicity.

It is natural to parameterize $Z_t$ as the negative force of a certain flow action $\tilde{S}$, i.e.
\begin{align}\label{eq:force}
    [Z^a(U_t)] (x, \mu) = - \partial^a_{x, \mu} \tilde{S}(U_t) \,,
\end{align}
where we define
\begin{align}
    \partial^a_{x, \mu} f(U) = \left. \frac{d}{d \tau} f(U_\tau) \right|_{\tau = 0}
    \label{eq:def_group_deriv}
\end{align}
with 
\begin{align}
     U_\tau (y, \nu) = \begin{cases}
    e^{\tau T^a} U(x, \mu), & \text{for } (x, \mu) = (y, \nu) \,, \\
    U(y, \nu) & \text{else} \,.
  \end{cases}
\end{align}
It can be shown \cite{luscher2010trivializing} that the determinant of the Jacobian of the redefinition $\mathcal{F}(V)=U_t$ is given by
\begin{align}
    \ln \det \mathcal{F}_* (V) = \int_0^t \textrm{d}s \, \mathcal{L}_0 \tilde{S}(U_s) \,,
\end{align}
with the Laplacian 
\begin{align}
    \mathcal{L}_0 = - \sum_{x,\mu} \partial^a_{x, \mu} \partial^a_{x, \mu} \,. \label{eq:laplacian}
\end{align}
In L\"uscher's approach, the flow action $\tilde{S}$ is given by a linear combination of a certain set of Wilson loops $\mathcal{W}_i$
\begin{align}\label{eq:flow_action}
    \tilde{S}(U_t, t) = \sum_{i} c_i(t) \, \mathcal{W}_i (U_t) \,,
\end{align}
where $c_i$ are coefficient functions that need to be determined. 
L\"uscher then proposed expanding this action in flow time perturbatively
\begin{align}\label{eq:t_expansion}
    \tilde{S}(U_t, t) = \sum_{k=0}^\infty t^k \tilde{S}^{(k)} 
\end{align}
with $t\in[0,1]$. L\"uscher explicitly determined the leading order $\tilde{S}^{(0)}$ and next-to-leading order $\tilde{S}^{(1)}$ of the expansion.
\\

\paragraph{Machine Learning Approach.}
We closely follow L\"uscher's construction but circumvent the perturbative approximation of the coefficients by using machine learning techniques. More specifically, we parameterize the coefficient functions by a simple ansatz, such as affine linear functions or cubic splines, and then learn their parameters $\theta$ by stochastic gradient descent. An advantage of this approach is that the free parameters of the coefficient functions can be initialized such that the coefficient functions match the perturbative results obtained by L\"uscher. In stark contrast to standard deep learning approaches, this provides a rigorous initialization scheme based on perturbation theory that can systematically be improved by incorporating perturbative corrections of higher order. In addition, the resulting approach only requires common tools in lattice gauge theories, such as the ability to compute a generic action $\Tilde{S}$ made of Wilson loops \eqref{eq:flow_action} and its force \eqref{eq:force}. Furthermore, only a very limited number of free parameters are required for the coefficient functions, resulting in a drastic overall reduction of the necessary number of parameters for an expressive model. This is beneficial both for training speed and interpretability, e.g. identifying Wilson loops that are most important for the trivialization.

The main technical challenge of such an approach is to calculate the gradients of the objective $\frac{\partial \mathcal{C}}{\partial \theta}$. This is non-trivial, as one has to take the parameter dependence of the flow equation \eqref{eq:ode} into account. We overcome this challenge by using the adjoint state method, see \cite{sengupta2014efficient} for a review. To this end, 
we derive a specific version of the adjoint state method for the Lie group $SU(N)$. Unlike previous work on the adjoint state method on manifolds \cite{falorsi2021continuous, mathieu2020riemannian, katsman2021equivariant, lou2020neural}, our method is particularly suited to the $SU(N)$ case and can be efficiently implemented using existing libraries for lattice field theory. 
We refer to the Supplementary Material
for a detailed derivation but summarize the main results in the following.

The adjoint state method starts from the observation that the optimization criterion \eqref{eq:kl} is to be minimized on the solution space of the differential equation \eqref{eq:ode}. 
We, therefore, introduce an $\mathfrak{su}$(N)-valued Lagrange multiplier $\lambda$, which in this context is also called the adjoint state, and define the Lagrangian 
\begin{align}
    L(\theta) = \mathcal{C}(\theta) - \left\langle \int_0^t \, \textrm{d} s \, \left( \lambda_s, \,\dot{U}_s U^\dagger_s - Z_s \right) \right \rangle_q \,,
\end{align}
with the standard inner product on the $\mathfrak{su}(N)$ Lie algebra
\begin{equation}
    ( A, B ) \equiv -2\sum_{x,\mu} \mathrm{tr} (A(x, \mu)B(x, \mu)) \,. 
\end{equation}
On the solution space, the objective $\mathcal{C}$ and the Lagrangian $L$ agree. By differentiating the Lagrangian, it can be shown that its gradient is given by
\begin{align}
    \frac{\partial \mathcal{C}}{\partial \theta} = \frac{\partial L}{\partial \theta} = \left \langle \int_0^t \textrm{d}s \, \left\{\left( \lambda_s, \partial_\theta Z_s \right) - \partial_\theta \mathcal{L}_0 \tilde{S}_s \right\}\right \rangle_q \label{eq:adjoint_state_gradient_main}
\end{align}
when the adjoint state fulfills the terminal value problem
\begin{align}
    &\dot{\lambda}_s = \partial \mathcal{L}_0 \tilde{S}_s + \left[ Z_s, \lambda_s \right] - \sum_{y, \nu} \lambda^a_s(y, \nu) \, \partial Z^a_s(y, \nu) \,, \nonumber \\
    &\lambda_t = \partial S(U_t) \,, \label{eq:ode_adjoint_state}
\end{align}
where we have used the notation $[\partial f(U)](x, \mu) = T^a \partial^a_{x, \mu} f(U)$, $\mathcal{L}_0$ is the Laplacian defined in \eqref{eq:laplacian}, and $t$ denotes the terminal flow time. 

Therefore, we can calculate the gradient $\partial_\theta \mathcal{C}$ by evolving the flow equation \eqref{eq:ode_adjoint_state} for the adjoint state $\lambda_s$ backwards in flow time. This has a comparable numerical cost to solving the flow equation \eqref{eq:ode} for the gauge configuration $U_t$. As a result, the cost of the adjoint state method does not scale with the number of parameters in stark contrast to finite differences.

Furthermore, it can be shown that the adjoint state $\lambda_0$ corresponds to the force of the action \eqref{eq:zeroflowtimeaction} at zero flow time 
\begin{align}
    \lambda^a_0(\mu, x) = \hat{\partial}^a_{\mu, x} S_{\mathcal{F}}(V) \,.
\end{align}
We note that this force can be used for a Hybrid Monte Carlo algorithm in the trivialized field $V$. We plan to explore this possibility as part of future work.
\\

\paragraph{Numerical Results.}
We compare our method to the state-of-the-art deep-learning approach of \cite{boyda2021sampling} which we estimate uses $\mathcal{O}(10^6)$ parameters. In order to compare, we closely follow their reported experiments. To this end, we consider the two-dimensional $SU(3)$ Yang-Mills theory with the standard Wilson action
\begin{equation}\label{eq:wilson_action}
    S_W(U) = -\frac{\beta}{6} \mathcal{W}_0(U),
\end{equation}
where $\mathcal{W}_0$ denotes the sum of plaquettes.
We train on a 16$\times$16 lattice size with $\beta$$=$$4.0$, $5.0$, and $6.0$ using two architectures: models A and B. Model A has 14 free parameters and uses affine linear coefficient functions for the seven Wilson loops of L\"uscher's perturbative construction. Model B has 420 free parameters and uses cubic splines with 10 knots as coefficient functions for 42 Wilson loops. Namely, all loops up to length 8 in combination with their moments and correlation functions of plaquettes are included in its flow action \eqref{eq:flow_action}.
We refer to the Supplementary Material for a detailed description. 

The training of model A is initialized using L\"uscher's perturbative solution. Since the flow action is a simple linear combination of Wilson loops, we initialize the training of model B from the trained model A. Specifically, we initialize the more expressive cubic spline coefficient functions such that they reproduce the affine linear coefficient functions learnt by the smaller model A and initialize the coefficients of the additional Wilson loops by small random numbers. This possibility for progressive training is a notable advantage of our approach. 

For time integration, we use 20 steps  of a third-order Crouch-Grossmann integrator \cite{crouch1993numerical} which is a suitable Runge-Kutta quadrature scheme for a Lie group. The Adam optimizer~\cite{kingma2014adam} with a mini-batch of size of 1024 and a learning rate of 0.0005 is chosen. Due to the low memory footprint of the model, each integration step can be checkpointed to reduce the error in the backward integration. The variance of the gradients is reduced by using the path-gradient VarGrad estimator \cite{richter2020vargrad, vaitl2022path, vaitl2022gradients}.
The quality of the model is quantified using the effective sampling size (ESS)
\begin{align}
    \textrm{ESS} = \frac{1}{\langle w(V)^2 \rangle_q} \in [0, 1] \,.
\end{align}
with the values reported in \cite{boyda2021sampling} for comparison.

\begin{table}[htb!]
    \centering
    \begin{tabular}{|lc|c|c|c|c|}
        \hline
        \multicolumn{2}{|c|}{\multirow{2}{*}{Ref.}} & \multirow{2}{*}{$N_{\rm params}$} & \multicolumn{3}{c|}{ESS at $\beta$} \\
        \multicolumn{2}{|c|}{} &  & $4.0$ & $5.0$ & $6.0$ \\
        \hline
        \hline
        ~L\"uscher, NL & \cite{luscher2010trivializing} & ~8 non-zero values~ & ~42\%~ & ~4\%~ & ~$<$1\%~
        \\
        \hline
        \hline
        \multirow{2}{*}{\textbf{This work}} & \textbf{A} & 14 $\equiv 2_t\times 7_W$ & 91\% & 65\% & 26\% 
        \\
         & \textbf{B} &~420 $\equiv10_t\times 42_W$~& 98\% & 88\% & 70\% 
        \\
        \hline
        \hline
        ~Boyda \textit{et al.} & \cite{boyda2021sampling} & $\mathcal{O}(10^6)$ estimated & ~88\%~ & ~75\%~ & ~48\%~
        \\
        \hline
    \end{tabular}
    \caption{For L\"uscher, the coefficients of the next-to-leading order calculations of \cite{luscher2010trivializing} are used, the ESS for Boyda \textit{et al.} is as reported in \cite{boyda2021sampling}. The total number of parameters $N_{\rm params} = N_t \times N_W$ of our approach is divided due to the number of Wilson loops $N_W$ and parameters per coefficient function $N_t$.}
    \label{tab:results}
\end{table}

\begin{figure}[htb]
    \centering
    \includegraphics[width=\linewidth]{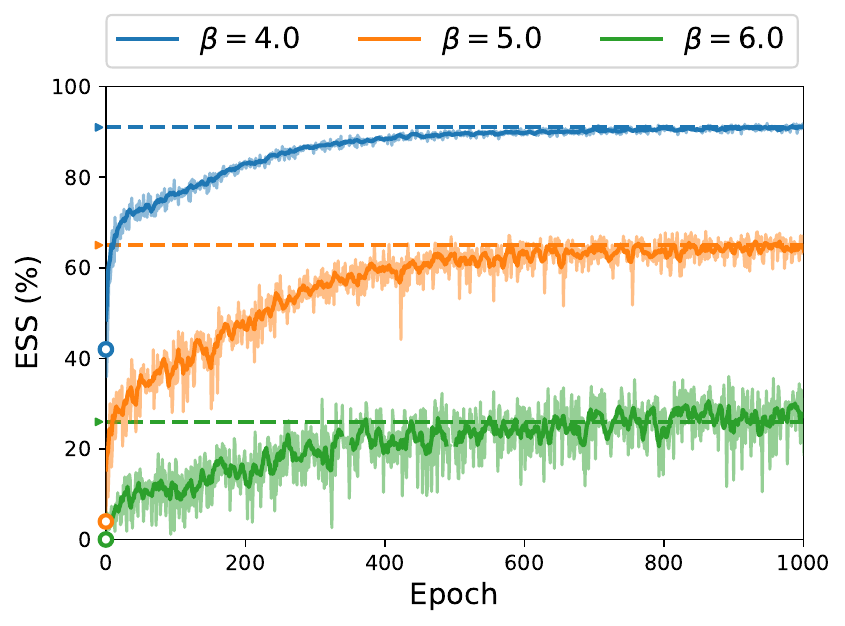}
    \caption{ESS measured during training of model A starting from L\"uscher's initialization. The faint line is the ESS over a single mini-batch. The thick line is a moving average over 6 steps. The empty circle at zero indicates the initial ESS. The horizontal dashed line is the ESS measured at high accuracy after training.}
    \label{fig:training}
\end{figure}

The results in Table~\ref{tab:results} show that our model B can significantly outperform the deep-learning-based approach by \cite{boyda2021sampling}. This is despite the fact that it has several orders of magnitude less parameters. This point is further illustrated by the fact that the smaller model A achieves comparable performance with only 14 parameters. Table~\ref{tab:results} also shows that our models lead to a significantly larger effective sampling size than the perturbative construction by L\"uscher, establishing the idea that machine learning can substantially improve upon the perturbative scheme.
At the same time, our approach can benefit from this perturbative scheme as the L\"uscher initialization provides a good starting point for training, see Figure~\ref{fig:training}. 

\begin{figure}[htb]
    \centering
    \includegraphics[width=0.9\linewidth]{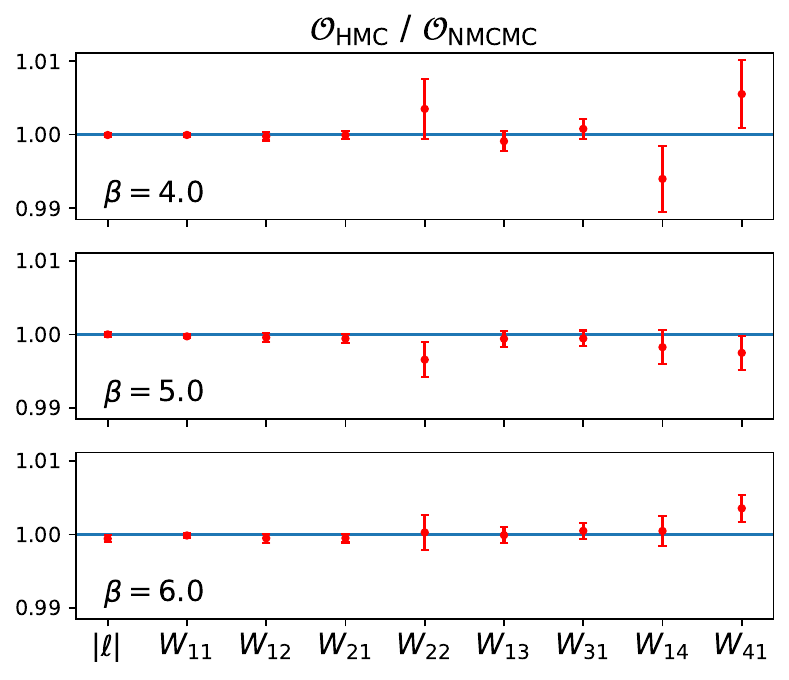}
    \caption{Observables estimated by both HMC and flow NMCMC~\cite{nicoli2020asymptotically}, see Eq.~\eqref{eq:NMCMC}, using a batch-size of 1024 configurations and 1000 samples, i.e. 1.024.000 samples in total.}
    \label{fig:comparison}
\end{figure}

As a final consistency check, Figure~\ref{fig:comparison} demonstrates the compatibility of estimates obtained by our method with ones from the HMC algorithm using the same observables as in \cite{boyda2021sampling}.
\\

\paragraph{Conclusion.} 
In this work, we have proposed a new method to learn trivializing maps which is natural for Lie groups and has several specific strengths: i) considerable parameter efficiency, ii) a high level of interpretability, iii) initialization based on perturbation theory, iv) equivariance with respect to all symmetries of the theory, v) progressive training, as well as vi) the possibility of implementation and parallelization with standard Lattice QCD libraries.   

To the best of our knowledge, the present work is the first to apply the adjoint state method in the context of lattice gauge theory.
This has the beneficial feature of providing the force of the flowed action and thus allows for HMC in the trivialized field variables $V$, as originally proposed in~\cite{luscher2010trivializing} and recently applied to normalizing flows in~\cite{albandea2022learning}. This constitutes a promising route for future research, in particular, as such a HMC could also be used during training. Our approach could also be applied to correct mistuned simulation parameters as an alternative to reweighting.
Furthermore, it can be naturally combined with domain decomposition algorithms, see \cite{luscher2005schwarz, finkenrath2022tackling}.

As the purpose of this paper was to benchmark our approach with respect to existing methods, we leave the application to the four-dimensional case, which manifestly suffers from critical slowing down \cite{schaefer2011critical}, for upcoming work. The beneficial properties of our method, as demonstrated by this work, make successful applications in this higher dimensional case significantly more plausible.  If successful, applications of machine-learning-based trivializing maps to full QCD would be within reach.\\

\begin{acknowledgments}
\paragraph{Acknowledgments.} We wish to acknowledge Michael G\"unther and Kevin Schäfers for useful discussions on Runge-Kutta integrators for Lie groups; and Jacob Finkenrath for useful discussions during the implementation. S.B. is supported by the H2020 project PRACE 6-IP (grant agreement No. 82376) and the EuroCC project (grant agreement No. 951740). P.K. is supported by the German Ministry for Education and Research (BMBF) as BIFOLD - Berlin Institute for the Foundations of Learning and
Data under grants 01IS18025A and 01IS18037A. S.S. is supported by the German Research Foundation (DFG) research unit FOR5269 ``Future methods for studying confined gluons in QCD''. L.V. thanks the Cyprus Institute for hospitality. Numerical results were obtained using the Cyclamen cluster of The Cyprus Institute equipped with P100 GPUs. The software is implemented using the Lyncs-API~\cite{Bacchio:2022bjk}, its interface to QUDA~\cite{Yamamoto:2022ygt} and the QUDA library~\cite{Clark:2009wm}. All figures were obtained by the authors under a CC BY 4.0 license\footnote{https://creativecommons.org/licenses/by/4.0/}, unless otherwise stated.
\end{acknowledgments}

\bibliography{main}
\nopagebreak
\appendix
{\hfill\textbf{SUPPLEMENTAL MATERIAL}\hfill}

\section{Relation between Objective and Trivializing Condition}\label{app:objective_trivialization_cond}

\begin{theorem}
The parameter value $\hat{\theta}$ is the global minimizer of the objective function 
\begin{align}
    \mathcal{C}(\theta) &= \langle S_{\mathcal{F}_\theta}(V) \rangle_q \,, \label{eq:app_objective_fn}
\end{align}
if and only if it fulfills the (partial) trivializing condition   
\begin{align}
    S_{\mathcal{F}_{\hat{\theta}}}(V) = S'(V) + \textrm{const.} \,,  \label{eq:app_trivializing_cond} 
\end{align}
where $S'$ is the action of the base density 
\begin{align}
    q(V) = \frac{1}{\mathcal{Z}'} \exp(-S'(V)) \,.
\end{align}
\end{theorem}
\begin{remark}
Note that for the choice $S'(V)=0$, the base density is uniform, i.e. $q(V)=\textrm{const.}$, and thus reduces to the case considered in the main text.
\end{remark}
\begin{proof}
We recall the map $\mathcal{F}_{\theta}$ induces a probability density 
\begin{align}
p_\theta(V) = \frac{1}{\mathcal{Z}} \exp(-S_{\mathcal{F}_\theta}(V)) 
\end{align}
by pullback of the target density $p(U)=\frac{1}{\mathcal{Z}} \exp(-S(U))$ of the lattice field theory. We also recall that its action $S_{\mathcal{F}_\theta}$ was defined by \eqref{eq:zeroflowtimeaction}.
Using this pullback density, the trivializing condition \eqref{eq:app_trivializing_cond} can then be rewritten as 
\begin{align}
    \log p_\theta(V) = \log q(V)  \,.  \label{eq:rewritten_tivializing_cond}
\end{align}
up to an irrelevant constant. This implies that the trivializing condition \eqref{eq:app_trivializing_cond} holds if and only if the two densities agree, i.e. $q \equiv p_\theta$.

On the other hand, the cost function can be rewritten as follows
\begin{align}
    \mathcal{C}(\theta) &= \langle S_{\mathcal{F}_\theta}(V) \rangle_q = \underbrace{\left\langle \log \frac{q(V)}{p_\theta(V)} \right\rangle_q}_{= \textrm{KL}(q, p_\theta)}  + \textrm{const.} 
    \label{eq:information_rewrite} \,,
\end{align}
where we have used that $q(V)$ is constant with respect to the parameters $\theta$ and the definition of the Kullback--Leibler (KL) divergence 
$\textrm{KL}(q, p_\theta) = \left\langle \log \tfrac{q(V)}{p_\theta(V)} \right\rangle_q$. 

It can be shown \citee{thomas2006elements} that the KL divergence is minimized if and only if the two densities are identical, $q \equiv p_\theta$. By \eqref{eq:information_rewrite}, the objective $C(\theta)$ is thus also minimized if and only if the two densities agree, i.e. if and only if the trivializing condition \eqref{eq:app_trivializing_cond} holds.
\end{proof}

We stress that the proof would work for any loss function corresponding to a probability divergence (and not only \eqref{eq:app_trivializing_cond} corresponding to the KL divergence) since one of the defining properties of a divergence is that it is globally minimized if and only if the two densities are the same.

\section{Adjoint State Method}\label{app:adjoint_state}
For training, we need to compute the gradient of the objective function \eqref{eq:app_objective_fn} with respect to the parameters $\theta$. This is challenging as we need to take the dependence on the parameters of the gauge fields into account
\begin{align}
    \frac{d}{d \theta_i} U_s(x, \mu) = Y^{(i)}_s(x, \mu) U_s(x, \mu)\,. \label{eq:app_yode}
\end{align}
As explained in the main text, we need to calculate the gradient on solutions of the flow equation \eqref{eq:ode} and therefore consider the Lagrangian 
\begin{align}
    L(\theta) = \mathcal{C}(\theta) - \left\langle \int_0^t \, \textrm{d} s \, \left( \lambda_s, \,\dot{U}_s U^\dagger_s - Z_s \right) \right\rangle_q \,,
\end{align}
The Lagrangian $L$ and the objective $\mathcal{C}$ agree on the solution space. We can thus also minimize the Lagrangian whose gradient is given by
\begin{align}
\frac{d L}{d \theta} = \langle \mathcal{G}_1 - \mathcal{G}_2 \rangle_q \,,
\end{align}
where 
\begin{align*}
\langle \mathcal{G}_1 \rangle_q &= \frac{d \mathcal{C}}{d\theta} = \left\langle \frac{d}{d\theta} S(U_t)  -  \frac{d}{d\theta} \int_0^t \textrm{d}s \, \mathcal{L}_0 \tilde{S}(U_s)  \right\rangle_q \,, \\ \langle \mathcal{G}_2 \rangle_q &= \frac{d}{d \theta} (\mathcal{C} - L) = \left\langle \frac{d}{d \theta}\int_0^t \, \textrm{d} s \, \left( \lambda_s, \,\dot{U}_s U^\dagger_s - Z_s \right) \right\rangle_q \,.
\end{align*}
The components of the first term $\mathcal{G}_1$ are easy to calculate
\begin{align}
    \mathcal{G}_1^{(i)} = \left( \partial S_t, Y^{(i)}_t \right) - \int_0^t \textrm{d}s \, \left( \partial \mathcal{L}_0 \tilde{S}_s , Y^{(i)}_s \right)- \int_0^t \textrm{d}s \, \partial_{\theta_i} \mathcal{L}_0 \tilde{S}_s \,.
\end{align}
To calculate the second term, we first note that
\begin{align}
    \frac{d}{d\theta_i} \dot{U}_s U_s^\dagger = \dot{Y}^{(i)}_s + [Y_s^{(i)}, Z_s] \equiv D_{Z_s} Y^{(i)}_s \,.  \label{eq:app_adj_deriv}
\end{align}
Using this result, the components of the second term are given by
\begin{align}
    \mathcal{G}^{(i)}_2 = \int_0^t \textrm{d}s \, \left( \lambda_s, \; D_{Z_s} Y^{(i)}_s - \partial_{\theta_i} Z_s -  \left( Y^{(i)}_s, \partial \right) Z_s \right)\,.
\end{align}
We recall that the Lie-algebra-valued fields $Y^{(i)}_s$ encode the dependency of the gauge fields $U_s$ on the parameter $\theta_i$. Since this dependency is non-trivial, we will choose the adjoint state $\lambda_s$ such that all contributions involving $Y^{(i)}_s$ vanish. For this, we need to collect all terms involving them. For the $\mathcal{G}_1$ term, this is straightforward. For the $\mathcal{G}_2$ term however, we use partial integration
\begin{align*}
\int_0^t \textrm{d}s &\, \left( \lambda_s, D_{Z_s} Y^{(i)}_s \right) = \\
    &\left(\lambda_t, Y^{(i)}_t \right) - \left( \lambda_0, Y^{(i)}_0 \right)- \int_0^t \textrm{d}s \, \left( D_{Z_s} \lambda_s, Y^{(i)}_s \right)
\end{align*}
and the identity
\begin{align*}
    \int_0^t \textrm{d}s &\, \left( \lambda_s, \; \left( Y^{(i)}_s, \partial \right) Z_s \right) = \\
    &\int_0^t \textrm{d}s \, \left( Y^{(i)}_s, \; \sum_{y, \nu} \lambda^a_s(y, \nu) \, \partial Z^a_s(y, \nu) \right)
\end{align*}
Using these expressions, we can rewrite the components $\mathcal{G}^{(i)}_2$ as
\begin{align*}
    \mathcal{G}^{(i)}_2 = -\int_0^t & \textrm{d}s \, 
     \left(Y^{(i)}_s , \; \; D_{Z_s} \lambda_s + \sum_{y, \nu} \lambda^a_s(y, \nu) \, \partial Z^a_s(y, \nu) \right) \\
     & - \int_0^t \textrm{d}s \, \left( \lambda_s, \;  \partial_{\theta_i} Z_s \right) + \left(\lambda_t, \; Y^{(i)}_t \right) - \left( \lambda_0, \; Y^{(i)}_0 \right) 
\end{align*}

Combining the two contributions to the gradient, we obtain
\begin{align*}
    \mathcal{G}^{(i)}_1 &- \mathcal{G}^{(i)}_2 = \\
    &\int_0^t  \textrm{d}s \, \left(Y^{(i)}_s ,  D_{Z_s} \lambda_s + \sum_{y, \nu} \lambda^a_s(y, \nu) \, \partial Z^a_s(y, \nu) - \partial \mathcal{L}_0 \tilde{S}_s \right) \\
    &+ \left(\partial S_t - \lambda_t, \; Y^{(i)}_t \right) \\
    & + \int_0^t \textrm{d}s \,\left \{ \left( \lambda_s, \;  \partial_{\theta_i} Z_s \right) - \partial_{\theta_i} \mathcal{L}_0 \tilde{S}_s \right \}  + \left( \lambda_0, \; Y^{(i)}_0 \right) \,.
\end{align*}
We can therefore avoid any terms involving the fields $Y^{(i)}_s$ by choosing the adjoint state $\lambda_s$ to be a solution of the following terminal value problem
\begin{align}
   & D_{Z_s} \lambda_s + \sum_{y, \nu} \lambda^a_s(y, \nu) \, \partial Z^a_s(y, \nu) - \partial \mathcal{L}_0 \tilde{S}_s = 0 \,, \\ 
   & \lambda_t = \partial S_t \,.
\end{align}
We note that this is the same differential equation as \eqref{eq:ode_adjoint_state} as can be seen by using the definition \eqref{eq:app_adj_deriv} of the adjoint covariant derivative $D_{Z_t}$. 

Provided that the adjoint state $\lambda_s$ is a solution to this terminal value problem, the gradient of the Lagrangian $L$ is then given by
\begin{align}
   \frac{\partial L }{ \partial \theta_i} & = \langle \mathcal{G}^{(i)}_1 - \mathcal{G}^{(i)}_2 \rangle_q \nonumber \\
   &= \left \langle \int_0^t \textrm{d}s \,\left \{ \left( \lambda_s, \;  \partial_{\theta_i} Z_s \right) - \partial_{\theta_i} \mathcal{L}_0 \tilde{S}_s \right \} + \left( \lambda_0, \; Y^{(i)}_0 \right)  \right \rangle_q \,. \label{eq:app_adjoint_der2}
\end{align}
The last term vanishes if $Y^{(i)}_0 = 0$, i.e. $U_0$ is independent of the parameters $\theta$. In this case, one recovers the equation for the adjoint state method \eqref{eq:adjoint_state_gradient_main} given in the main text.

\section{Relation between adjoint state and the force}
In the following, it is shown that the adjoint state can be interpreted as the force of the action $S_{\mathcal{F}}$ defined in \eqref{eq:zeroflowtimeaction}. In order to see this, we first choose the following slightly modified initial conditions
\begin{align}
    V(y,\nu) \to V_\tau(y,\nu) = 
      \begin{cases}
    e^{\tau T^a} V(x, \mu), & \text{for } (x, \mu) = (y, \nu) \,, \\
    V(y, \nu) & \text{else} \,,
  \end{cases} \label{eq:app_modifiedinit}
\end{align}
i.e. we multiply the $(x,\mu)$ component of the initial condition $V$ with the group element $e^{\tau T^a}$ and leave all other components unchanged. 
We stress that we do not make the dependency on the indices $a$, $\mu$, and $x$ explicit in order to simplify notation. 
The definition of the group derivative \eqref{eq:def_group_deriv} implies that the derivative with respect to $\tau$ of the operator $C$ corresponding to the cost function $\mathcal{C}=\langle C \rangle_q$  is 
\begin{align}
 \frac{d}{d\tau} \left. C(V_\tau) \right|_{\tau = 0}  = \hat{\partial}^a_{\mu, x} C(V) \,,
\end{align}
where the hatted derivative $\hat{\partial}$ is with respect to the initial field $V$.
The flow equation corresponding to the auxiliary parameter $\tau$ is given by
\begin{align}
    \frac{d}{d \tau} U_{\tau, s} = \mathcal{F}_{*, \tau, s} \, U_{\tau, s} \,,
\end{align}
where we have defined $U_{\tau, s} = \mathcal{F}_s(V_{\tau})$ and $\mathcal{F}_{*, \tau, s}$ is the algebra-valued generator field corresponding to the auxiliary parameter $\tau$, i.e. corresponds to $Y^{(i)}_s$ in the derivation of the adjoint state method for $\theta_i = \tau$, see \eqref{eq:app_yode}. 

From the definition of the modified initial conditions
\eqref{eq:app_modifiedinit}, it follows that
\begin{align}
\mathcal{F}^{b}_{*,0,0}(\nu, y ) = \left[ \left. \frac{d}{d \tau} V_\tau \right|_{\tau=0} V^\dagger \right]^b(\nu, y) =  \delta^{ab} \delta_{xy} \delta_{\mu \nu} \label{eq:app_explicit_F}
\end{align}

Since the field $Z_s$ and the Laplacian $\mathcal{L}_0 \tilde{S}_s$ do not explicitly depend on $\tau$, we obtain from the adjoint state method \eqref{eq:app_adjoint_der2} that\footnote{While we have derived the adjoint state method for the gradient of an \emph{expectation value} of an operator in the previous section, it can be easily seen from the derivation that it also holds for the operator itself.}
\begin{align}
    & \frac{d}{d\tau} \left. C(V_\tau) \right|_{\tau = 0}  = \hat{\partial}^a_{\mu, x} C(V)  =  (\lambda_{0, 0}, \mathcal{F}_{*,0,0} )\,, 
\end{align}
where $\lambda_{\tau,s}$ is the adjoint state corresponding to the modified initial condition \eqref{eq:app_modifiedinit}. In particular, this reduces to the adjoint state with respect to the original initial conditions $V$ if the auxiliary $\tau$ parameter is set to zero, i.e. $\lambda_{0,s} = \lambda_s$.
Using the explicit form of the differential \eqref{eq:app_explicit_F}, we thus conclude that
\begin{align}
 \hat{\partial}^a_{\mu, x} C(V) = \sum_{y, \nu} \lambda_0^b(\nu, y) \mathcal{F}^{b}_{*,0,0}(\nu, y) = \lambda_0^a (\mu, x) \,.
\end{align}
For the standard choice of the objective function \eqref{eq:app_objective_fn}, the adjoint state is then equal to the force with respect to the initial field $V$, i.e.
\begin{align}
    \lambda^a_0(\mu, x) = \hat{\partial}^a_{\mu, x} S_{\mathcal{F}}(V) \,.
\end{align}
We stress that this last step crucially relies on the choice of cost function and, therefore, may not hold for other choices.

\section{Detailed description of the flow models}

\begin{figure*}[htb!]
    \centering    \includegraphics[width=0.9\linewidth]{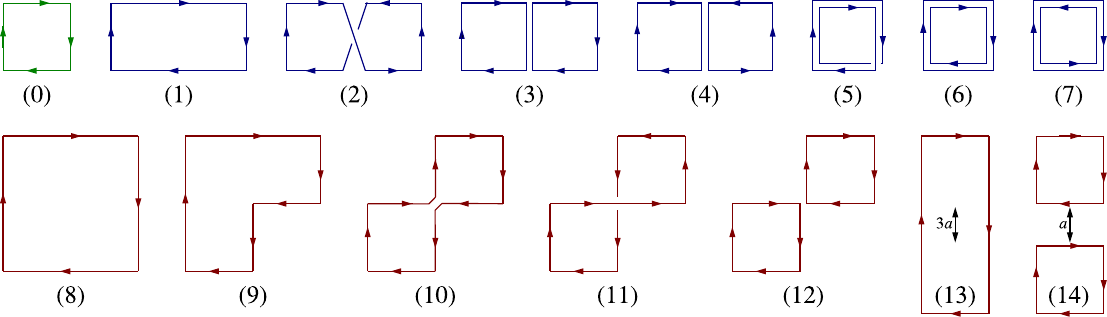}
    \caption{Classes of loops used in this work. We follow from 0-7 the same notation as~\cite{luscher2010trivializing}, from where the representation of the loops is borrowed. We use colors to separate categories of loops: in green we depict the plaquette that is used in the Wilson action $S_W$~\eqref{eq:wilson_action} and in the leading order of L\"uscher's expansion  $\tilde{S}^{(0)}$~\eqref{eq:t_expansion}; in blue we depict the loops in the next-to-leading order of L\"uscher's expansion $\tilde{S}^{(1)}$~\eqref{eq:t_expansion} -- model A is made by green and blue; and in red we depict the additional loops used in model B, namely all loops of length 8 and additional correlation of plaquettes. Model B includes also moments of all loops that do not have repeated links.}
    \label{fig:loops}
\end{figure*}

In Table~\ref{tab:results}, we have presented results for two models: A) with 14 parameters and B) with 420 parameters. In the following we give an in depth overview on how these models are constructed. We first introduce definitions and properties, and conclude the section with the description of the models.\\

\paragraph{Class of loops.} In Figure~\ref{fig:loops} we depict the class of loops that are used. We employ the same notation as in~\cite{luscher2010trivializing}, indicating with $\mathcal{W}_i$ the $i$th loop in the figure and defining
\begin{align*}
\mathcal{W}_i &= \sum_{C\in\Gamma_i} \tr\{U(C)\} &{\rm for~} i=0,1,2,5,8,9,10,11,13\\
\mathcal{W}_i &=\!\!\! \sum_{C,C'\in\Gamma_i}\!\!\! \tr\{U(C)\} \tr\{U(C')\} \!\!\!\!\!\!\!\!\!\!\!\!\!\!\!\!&{\rm for~} i=3,4,6,7,12,14
\end{align*}
where $U(C)$ denotes the ordered product of the link variables along the loop $C$. The
sums in these equations run over $\Gamma_i$, which includes all possible positions and orientations of loops in the $i$th class excluding repetitions. Since loops with opposite direction always appear, then all $\mathcal{W}_i$ are real-valued scalar quantities. The coefficients $c_i$ are unique for all loops $\Gamma_i$ achieving equivariance with respect to all lattice symmetries of the theory, namely discrete translation, rotation and orientation invariance.
\\

\paragraph{Mandelstam Constraints.}
From the Cayley-Hamilton (CH) theorem one can obtain identities for traces of $SU(N)$ matrices known as Mandelstam constraints~\citee{mandelstam1979charge,Boyle:2022xor}. 
For a $3\times3$ $SU(3)$ matrix the CH theorem reads as
\begin{equation}\label{eq:CH3}
U^2 = \tr(U) U -\tr(U)^\star \mathbb{I} + U^\dagger.
\end{equation}
By multiplying on both sides by a generic matrix $V$ and tracing, one obtains 
\begin{equation}
\tr(U^2V) = \tr(U)\tr(UV) - \tr(U)^\star\tr(V) + \tr(U^\dagger V).
\end{equation}
Special cases that follow are
\begin{align*}\label{eq:power_2}
    \tr(U^2) &= \tr(U)^2 - 2\tr(U)^\star\,, \qquad{\rm and}\\
    \tr(U^{n+1}) &= \tr(U) \tr(U^{n}) - \tr(U)^\star\tr(U^{n-1}) + \tr(U^{n-2})\,. \nonumber
\end{align*}
Using the latter, any trace of powers 
of loops can be written in terms of powers of traces of loops.

This is a critical step for a stable training, because certain Wilson loops introduce degeneracies, e.g.
\begin{equation}\label{eq:double_plaquette}
    \mathcal{W}_5 = \mathcal{W}_6 -2 \mathcal{W}_0,
\end{equation}
and they must be removed to achieve convergence during training. For this reason, in our models we choose to consider only powers of traces and not traces of powers.\\

\paragraph{Moments of loops.} 
Following the above discussion, we introduce moments of loops. We define
\begin{equation}\label{eq:moments}
\mathcal{W}_i^{(m,n)} = \sum_{C\in\Gamma_i} {\rm Re}(\tr\{U(C)\})^m~{\rm Im}(\tr\{U(C)\})^n,
\end{equation}
which is the $m$th-moment of the real part and the $n$th-moment of the imaginary part of a loop. Since the action is real, we consider only moments where $n$ is an even integer.
Additionally, we note that the following identities hold
\begin{align*}
     \mathcal{W}_i^{(1,0)} &= \mathcal{W}_i & {\rm for~} i=0,1,2,5,8,9,10,11,13\\
     \mathcal{W}_0^{(2,0)} &= 2\mathcal{W}_7+\mathcal{W}_6\\
     \mathcal{W}_0^{(0,2)} &= 2\mathcal{W}_7-\mathcal{W}_6.\\
\end{align*}

\paragraph{Correlation of loops.} Finally, the loops $\mathcal{W}_3$, $\mathcal{W}_4$, $\mathcal{W}_{12}$ and $\mathcal{W}_{14}$ belong to the class of correlation of loops. For generalization purpose, we define
\begin{align}\label{eq:correlation}
\mathcal{W}_{\langle i,j\rangle(\vec{d})}^{(m,n),(m',n')}  =
\!\!\!\!\!\!\!\!\!\!
\sum_{C,C'\in\Gamma_{\langle i,j\rangle(\vec{d})}}
\!\!\!\!\!\!\!\!\!
&\Big( {\rm Re}(\tr\{U(C)\})^m~{\rm Im}(\tr\{U(C)\})^n\nonumber\\
&\cdot{\rm Re}(\tr\{U(C')\})^{m'}{\rm Im}(\tr\{U(C')\})^{n'} \Big)
\end{align}
as the generic correlation of moments of two loops, where $\Gamma_{\langle ij\rangle(\vec{d})}$ runs over all possible pairs of loops $i$ and $j$ with separation $\vec{d}$. We note that 
\begin{align*}
    \mathcal{W}_{\langle 0,0\rangle(1)}^{(1,0),(1,0)} &= \mathcal{W}_4+\mathcal{W}_3\\
    \mathcal{W}_{\langle 0,0\rangle(1)}^{(0,1),(0,1)} &= \mathcal{W}_4-\mathcal{W}_3
\end{align*}
As for the moments of loops, since the action is real, we consider only correlations where $n+n'$ is an even number.\\

\begin{figure*}[htb!]
    \centering
    \includegraphics[width=0.95\linewidth]{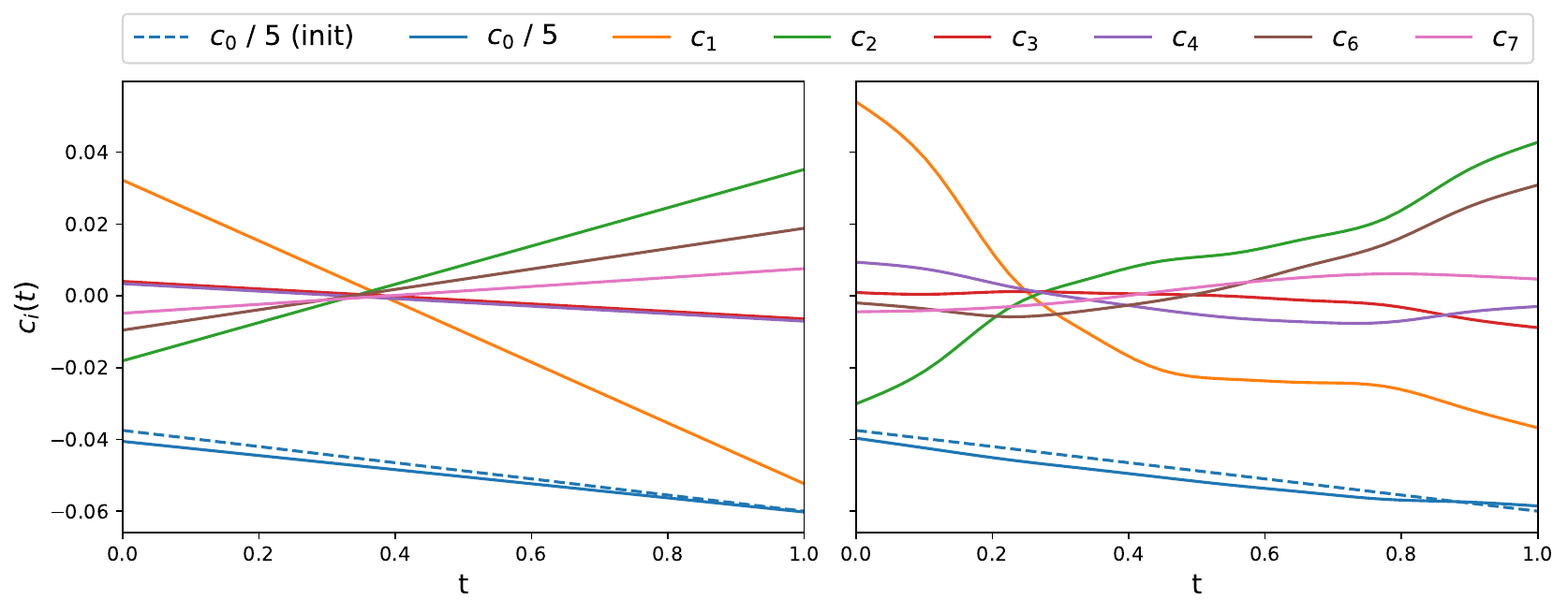}
    \caption{Comparison of the coefficient functions in model A (left panel) and their values in  model B (right panel), both trained at $\beta=6.0$. The coefficient functions of the plaquettes, $c_0$, are scaled by a factor of 0.2 and the dashed line shows the initial guess from L\"uscher's perturbative expansion.}
    \label{fig:coefficients}
\end{figure*}

\paragraph{Model A.}
The first model is inspired by the next-to-leading order of L\"uscher's perturbative expansion~\eqref{eq:t_expansion}. This is given by the terms $\tilde{S}^{(0)}$ and $\tilde{S}^{(1)}$, which have been computed in~\cite{luscher2010trivializing} for the Wilson action $S_W$~\eqref{eq:wilson_action}. At next-to-leading order the coefficient functions $c_i(t)$ are linear, where the intercept at $t=0$ is given by $\tilde{S}^{(0)}$ and the slope by $\tilde{S}^{(1)}$. For the Wilson action, $\tilde{S}^{(0)}$ is proportional to the plaquette only, $\mathcal{W}_0$, and therefore $c_0(t)$ is the only coefficient function having a non-zero intercept. $\tilde{S}^{(1)}$, instead, is proportional to those loops that are made by combining two plaquettes that share one link. These are $\mathcal{W}_1$ to $\mathcal{W}_7$. We note that applying the identity in~\eqref{eq:double_plaquette}, the slope of $\mathcal{W}_0$ is given by $\mathcal{W}_5$, which is excluded from the model to avoid degeneracy.
Therefore model A uses seven loops and two parameters for each coefficient function.

\paragraph{Model B}
The second model is an extension of Model A. It includes all moments~\eqref{eq:moments} that have $m+n\leq 3$, namely $(1,0)$, $(2,0)$, $(3,0)$, $(0,2)$ and $(1,2)$. These are computed for single Wilson loops up to length 8, namely $\mathcal{W}_0$, $\mathcal{W}_1$, $\mathcal{W}_8$, $\mathcal{W}_9$, $\mathcal{W}_{10}$, $\mathcal{W}_{11}$ and $\mathcal{W}_{13}$. These account for 35 terms. We did not include $\mathcal{W}_2$, which is also of length 8, but whose moments cannot be computed due to limitations of the implementation. Therefore only its real part is considered, namely $\mathcal{W}^{(1,0)}_2$. Then we include the correlation of plaquettes~\eqref{eq:correlation} at distance $1$, $(1,1)$ and $2$, namely $\mathcal{W}_{\langle 0,0\rangle(1)}^{(1,0),(1,0)}$, $\mathcal{W}_{\langle 0,0\rangle(1)}^{(0,1),(0,1)}$, $\mathcal{W}_{\langle 0,0\rangle(1,1)}^{(1,0),(1,0)}$, $\mathcal{W}_{\langle 0,0\rangle(1,1)}^{(0,1),(0,1)}$, $\mathcal{W}_{\langle 0,0\rangle(2)}^{(1,0),(1,0)}$ and $\mathcal{W}_{\langle 0,0\rangle(2)}^{(0,1),(0,1)}$.

The above sums up to 42 loops and, for each of these, a cubic spline interpolating 10 knots is used as a coefficient functions. In total Model B has 420 parameters. As depicted in Figure~\ref{fig:coefficients}, the coefficient functions can become highly nontrivial during training and we observe a significant improvement in the performance of the model up to 10 knots of the cubic spline. Further increasing the number of nodes has a less pronounced effect.

\section{Authorship}
As this is relevant for the PhD of L.V., the authors
want to state that they have used alphabetical ordering
for the list of authors as is conventional in the lattice
community. In the conventions of the machine learning
community, the contributions of L.V. and S.B. would be
equivalent to a joint first authorship.

\nopagebreak[4]
\bibliographystylee{unsrt}
\bibliographye{main}

\end{document}